\documentclass[draftclsnofoot,onecolumn]{IEEEtran}
\IEEEoverridecommandlockouts
\usepackage{cite}
\usepackage{amsmath,amssymb,amsfonts}
\usepackage{bm}
\usepackage{graphicx}
\usepackage{textcomp}
\usepackage{xcolor}
\usepackage[center]{caption}
\usepackage{amsthm}
\usepackage{bbm}

\usepackage{graphicx}
\usepackage{subfig}

\makeatletter
\newif\if@restonecol
\makeatother

\usepackage[linesnumbered,ruled,vlined]{algorithm2e}
\usepackage{algpseudocode}
\usepackage{amsmath}

\def\BibTeX{{\rm B\kern-.05em{\sc i\kern-.025em b}\kern-.08em
    T\kern-.1667em\lower.7ex\hbox{E}\kern-.125emX}}
\begin{document}

\title{UAVs as a Service: Boosting Edge Intelligence for Air-Ground Integrated Networks}

\author{Chao~Dong,
        Yun~Shen,
        Yuben~Qu,
        Qihui~Wu,
        Fan~Wu,
        and~Guihai~Chen
\thanks{C. Dong, Y. Shen, and Q. Wu are with the Key Laboratory of Dynamic Cognitive System of Electromagnetic Spectrum Space, Nanjing University of Aeronautics and Astronautics, China.}
\thanks{Y. Qu (\textit{Corresponding author}), F. Wu, and G. Chen are with the Department of Computer Science and Engineering, Shanghai Jiao Tong University, China.}
}

\maketitle

\begin{abstract}
The air-ground integrated network is a key component of future sixth generation (6G) networks to support seamless and near-instant super-connectivity. There is a pressing need to intelligently provision various services in 6G networks, which however is challenging. To meet this need, in this article, we propose a novel architecture called UaaS (\underline{U}AVs \underline{a}s \underline{a} \underline{S}ervice) for the air-ground integrated network, featuring UAV as a key enabler to boost edge intelligence with the help of machine learning (ML) techniques. We envision that the proposed UaaS architecture could intelligently provision wireless communication service, edge computing service, and edge caching service by a network of UAVs, making full use of UAVs' flexible deployment and diverse ML techniques. We also conduct a case study where UAVs participate in the model training of distributed ML among multiple terrestrial users, whose result shows that the model training is efficient with a negligible energy consumption of UAVs, compared to the flight energy consumption. Finally, we discuss the challenges and open research issues in the UaaS.
\end{abstract}

\section{Introduction}
While the fifth generation (5G) networks are being commercially available and deployed at 2020, both the academia and industry around the world have already paid attention to the research on the 5G beyond and sixth generation (6G) networks \cite{AIEI}. 6G networks substantially differ from previous generation networks in the revolution of realizing intelligence to satisfy more stringent demands and requirements for the expected intelligent information society by 2030 \cite{AIEI}, \textit{e.g.}, global 3D connectivity, massive connection ($10^7$ devices/km$^2$), very high user-experienced data rate ($\geq$ 1 Tb/s), very low latency (10-100 $\mu s$), computation-intensive and data-craving applications with low latency supporting (such as VR/AR/MR, autonomous driving, and SHD/EHD video streaming), etc. In a word, employing artificial intelligence (AI) techniques, how to intelligently provision the foregoing ubiquitous services in terms of communication, computing, and caching is a key to the success of 6G networks.


Meanwhile, it is recognized and expected that an integrated space-air-ground-underwater network will be the core potential architecture of future 6G networks, to support seamless and near-instant super-connectivity \cite{AIEI}. As a critical part therein, the integrated air-ground branch where unmanned aerial vehicles (UAVs) or drones are flexibly deployed in the air-network tier, is able to assist the communication, computing, and caching of the terrestrial network due to its flexibility and proximity \cite{AGIM}. Nevertheless, in the context of 6G networks, the air-ground integrated network is faced with several challenges, \textit{e.g.}, temporal-spatial dynamic communication/computing/caching service demand, large-scale complex connection decision and resource management, and ubiquitous intelligence demand inside the network. To summarize, it is extremely challenging to realize those visions of 6G in the air-ground integrated network.

There are several studies about the air-ground integrated network architecture in existing literatures. Cheng \textit{et al.} \cite{AGIM} propose a novel air-ground integrated mobile edge network, which investigates the potential benefits and applications of drone cells, and UAV-assisted edge computing and caching. To support diverse vehicular services, Zhang \textit{et al.} \cite{SDSA} propose a software defined networking (SDN)-based space-air-ground integrated network architecture. Focusing on provisioning computing services by UAVs, Zhou \textit{et al.} \cite{AAGI} propose an air-ground integrated mobile edge computing (MEC) framework to cater for the urgent computing service demand from internet of things (IoTs). Besides, Kato \textit{et al.} \cite{OSAG} present a comprehensive study about how to deal with the challenges related to the space-air-ground integrated networks by AI techniques, including network control, spectrum management, energy management, routing and handover management, and security guarantee. However, how the air-network tier intelligently provision the communication/computing/caching services for the ground-network tier in the air-ground integrated network remains unsolved so far.

In this article, we propose a novel \underline{U}AVs \underline{a}s \underline{a} \underline{S}ervice (UaaS) architecture that features UAV as a key enabler to boost edge intelligence for the air-ground integrated network. Specifically, we envision that UAVs can be as a wireless communication service, an edge computing service, or an edge caching service in the proposed UaaS, where all services are intelligently provisioned with machine learning (ML) techniques. In addition, each service can be provided by a network of multiple UAVs in a cooperative manner. For the first service, UAVs act as an ubiquitous aerial communication platform and an intelligent edge network regulator. For the second one, UAVs serve as a smart edge computing platform and an effective edge intelligence trainer. For the last one, UAVs play the roles of an intelligent content cache provider and a clever edge content sharing assistant.

From a high-level view, our solutions in the UaaS to address the aforementioned challenges are as follows. First, to satisfy the temporal-spatial dynamic service demand, we provision on-demand services by dispatching UAVs according to the prediction by ML techniques and big data analytics. Second, to combat with large-scale complex connection decision and resource management, we manage connections and resources by deep reinforcement learning (DRL) \cite{DRL,ADRLL}, benefited from the wide coverage of UAVs. Third, to meet the ubiquitous intelligence demand inside the network, we endow terrestrial users with ubiquitous powerful computation ability by UAVs, and enable UAVs to participate in the model training of distributed ML with the flexible deployment of UAVs.
\begin{figure*}[t]
\centering
\includegraphics[scale = 0.35]{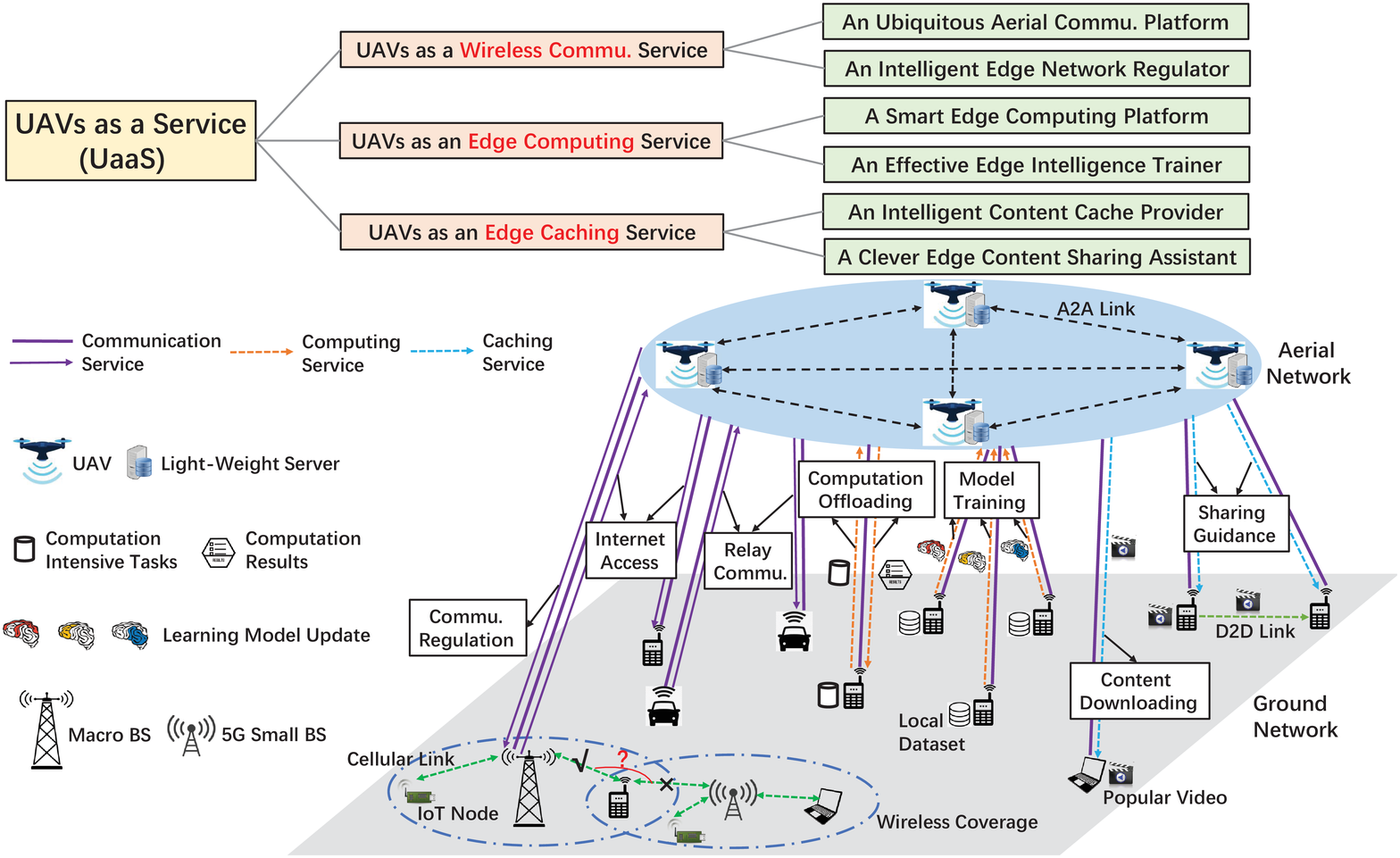}
\caption{An overview of the UaaS architecture.}
\label{Overview}
\vspace{-0.55cm}
\end{figure*}

Our proposed UaaS differs from the existing UAV-based air-ground integrated network architectures such as \cite{AGIM,AAGI,SDSA} in two main aspects. First, although some previous studies have expected that UAVs can provide communication/computing/caching service to the terrestrial network, most of them do not consider how to leverage existing AI technologies to intelligently provision these services. In contrast, we envision that those services can be provisioned by UAVs in an intelligent manner employing AI technologies, to boost edge intelligence expected by 6G. Second, compared to existing studies focusing on either UAV-enabled services or UAV-assisted services, the UaaS provides a comprehensive framework of various services related to UAVs.


In the rest of this article, we introduce the architecture overview of the proposed UaaS as well as the envisions, challenges, and solutions, explain the three services in detail respectively, conduct a case study of UAV-enabled federated learning, discuss the challenges, and highlight some promising future research directions.

%

\section{UAVs as a Service: Architecture, Envisions, Challenges, and Solutions}
In this section, we first describe the UaaS architecture with its envisions. Then, we briefly introduce the main challenges faced by the air-ground integrated network, with the corresponding solutions in the UaaS.
\subsection{Architecture Overview and Envisions}
The UaaS architecture is illustrated in Fig.~\ref{Overview}. In general, it has a two-layer air-ground networking architecture, where multiple flying UAVs in the air form the aerial network, and various mobile users and basic radio access networks (RANs) constitute the terrestrial network. In the upper aerial network, each UAV is equipped with wireless communication module, embedded computing processor (\textit{i.e.}, light-weight server), storage device, and various IoT devices (\textit{e.g.}, cameras, sensors, RFIDs, and actuators). In this sense, UAVs are expected to have considerable abilities of communication, computation, and storage, to provide a variety of aerial services in future 6G. In the lower terrestrial network, mobile users including mobile equipments (phones/notebooks), autonomous robotics/vehicles, and IoT devices, are served by the heterogeneous fundamental RANs consisting of WiFi, small cells, and macrocells, which can also be served by UAVs from the sky. UAVs are interconnected through aerial-to-aerial (A2A) communications such as WiFi, while the UAV aerial network and terrestrial network are connected via aerial-to-ground (A2G) and ground-to-aerial (G2A) communications.

We envision that the proposed UaaS architecture is able to intelligently provision various services by taking the advantage of UAVs and development of AI technologies. More specifically, we believe that UAVs can be perceived as an on-the-fly C$^3$ service in UaaS for future 6G networks as follows. For the first C, UAVs are as a wireless \underline{C}ommunication service by acting as an ubiquitous aerial communication platform and intelligent edge network regulator. For the second C, UAVs are as an edge \underline{C}omputing service by playing as a smart edge computing platform as well as an effective edge intelligence trainer. For the last C, UAVs are as an edge \underline{C}aching service by casting as a intelligent content cache provider and clever edge sharing assistant. We note that the latter two services are derived from the communication service, as any service provisioned by UAVs must be on the premise of necessary A2G or G2A communication links, as shown in Fig.~\ref{Overview}.
\begin{figure*}[t]
\centering
\includegraphics[scale = 0.425]{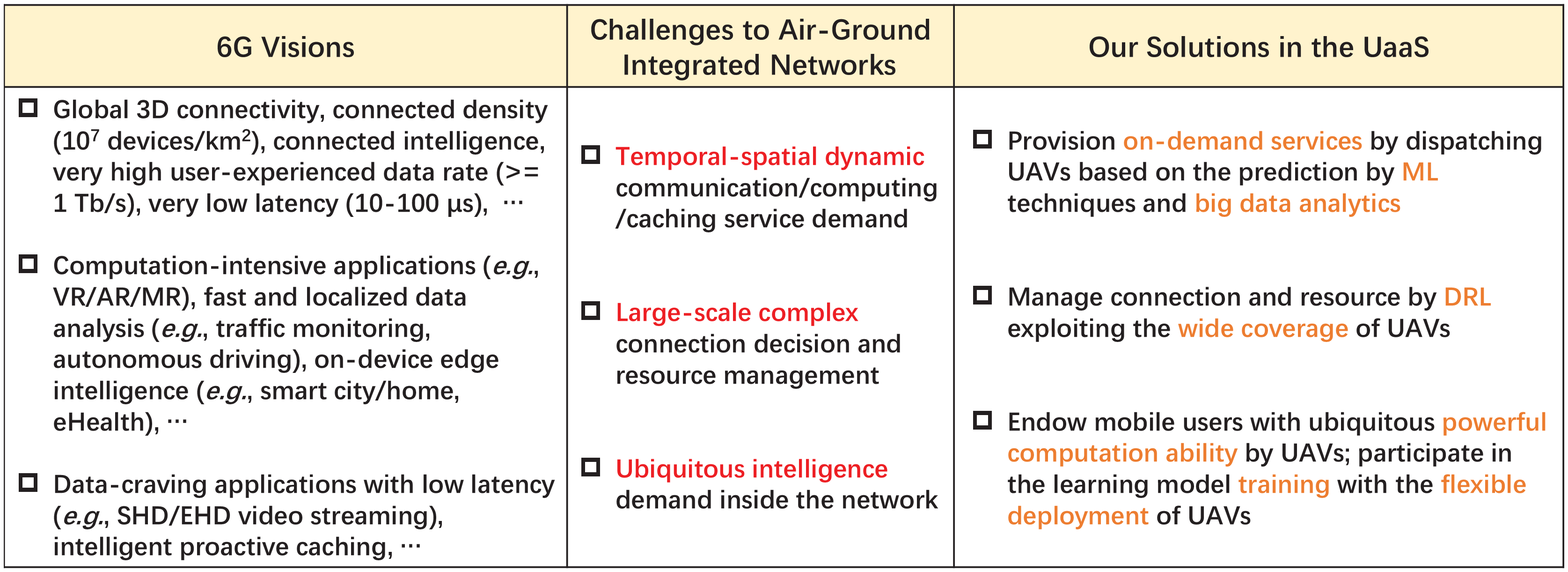}
\caption{6G visions, challenges, and our solutions in the UaaS architecture.}
\label{Challenges}
\vspace{-0.5cm}
\end{figure*}
\vspace{-0.35cm}
\subsection{Challenges and Solutions}
Due to the visions of future 6G networks as illustrated in Fig.~\ref{Challenges}, there exist three main challenges in the air-ground integrated networks as follows. First, the service demands from communication, computing, and caching are dynamic in the dimensions of both temporal and spatial. Second, it usually needs to address the problem of connection decision and resource management, \textit{e.g.}, access and handover decision, spectrum, computing, and storage resource allocation, which is generally large-scale and complex and hard to solve. Last, since intelligence is expected to be inherent in 6G networks, there is also an ubiquitous intelligence demand inside the network. We propose the corresponding solutions in the UaaS as follows. First, based on the prediction by ML techniques and big data analytics, we provision on-demand services by dispatching UAVs to adaptively cater for the demand. Second, exploiting the wide coverage of UAVs as well as the power of DRL in dealing with large-scale complex problems, we manage the massive connections and various resources by DRL. Last, to intelligentize the network, we endow mobile users with ubiquitous powerful computation ability by UAVs, and enable UAVs to participate in the distributed ML model training with the flexible deployment of UAVs.

\section{UAVs as a Wireless Communication Service}
We first introduce how UAVs can be as an appealing wireless communication service to support the huge and highly dynamic data traffic. Although the communication benefits of UAVs have been well recognized in existing works such as \cite{AFTS}, how those benefits are intelligently realized with the help of ML techniques is still in its infancy, which is however the main focus of this paper.

\begin{figure*}[t]
\centering
\subfloat[UAVs as an ubiquitous aerial communication platform]{\includegraphics[height=2in]{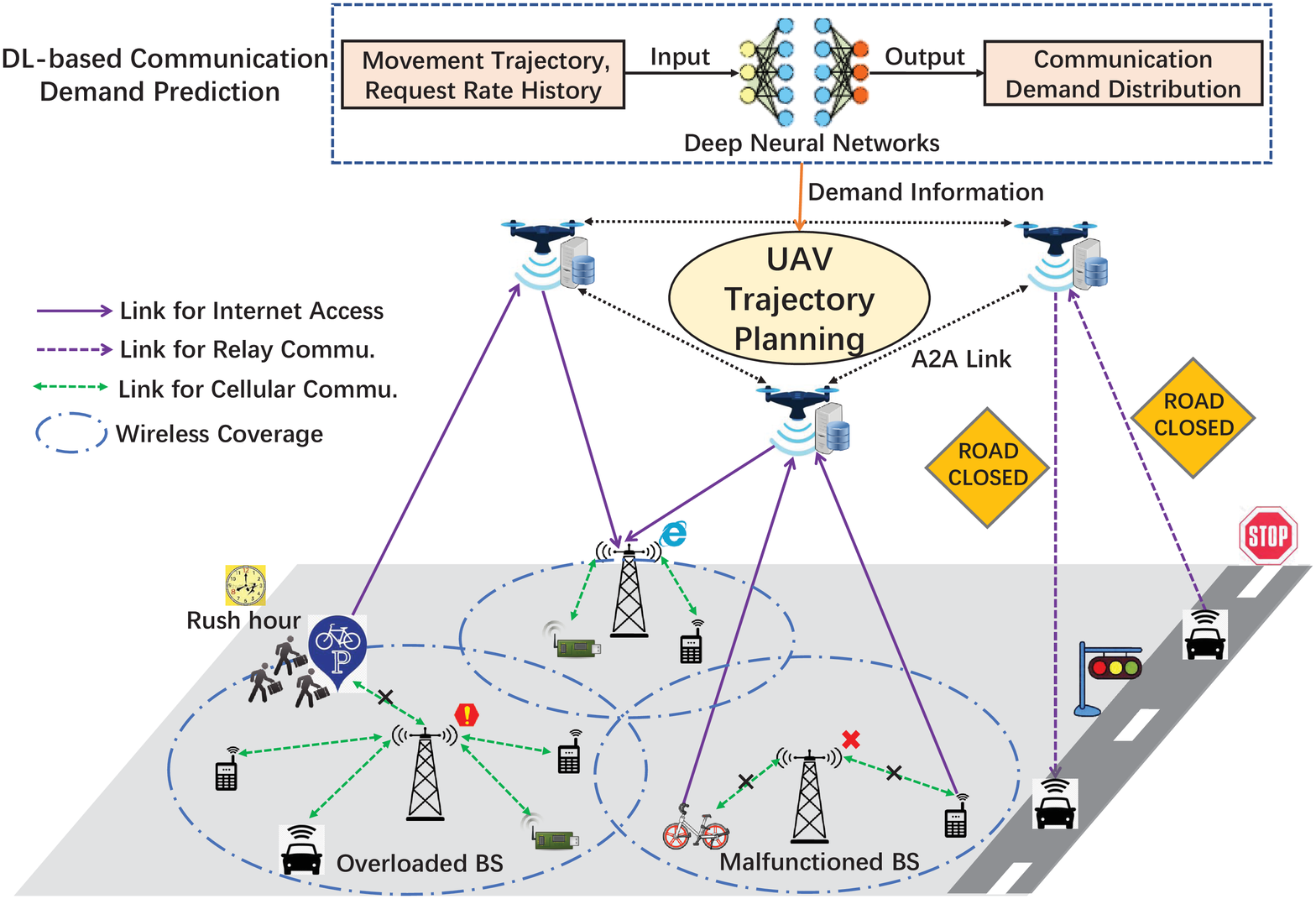}}\hspace{0.3cm}
\subfloat[UAVs as an intelligent edge network regulator]{\includegraphics[height=2in]{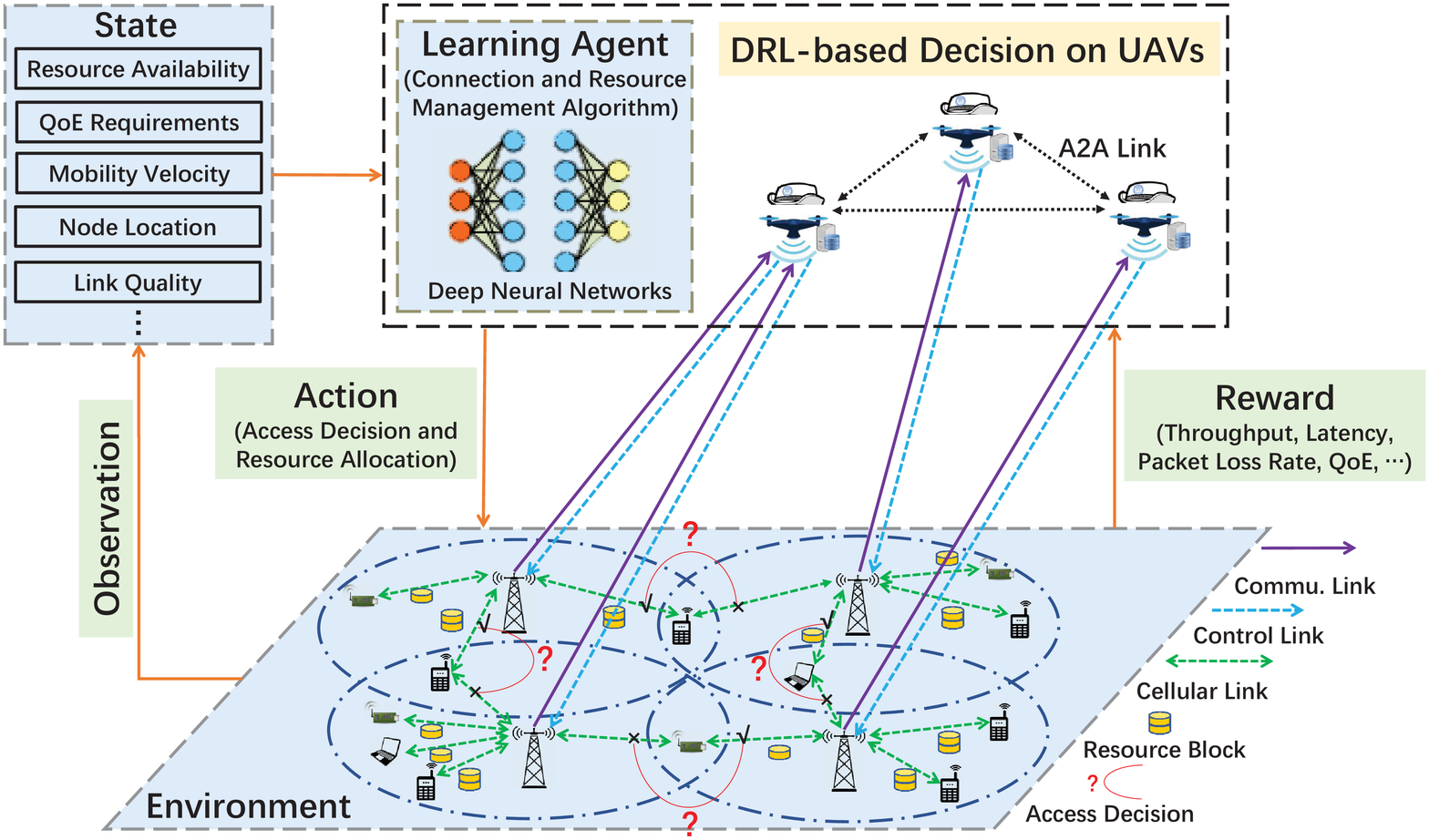}}\hfill
\caption{UAVs as a Wireless Communication Service.}
\label{UaaS_C1}
\vspace{-0.5cm}
\end{figure*}

\textbf{UAVs as an ubiquitous aerial communication platform:} UAVs are employed to serve as aerial APs or relays for mobile users and IoT devices. Although the terrestrial networks including 5G are able to achieve pretty high data rates, the network coverage in some remote areas such as rural regions is poor, and existing deployed terrestrial networks are hard to serve the highly dynamic mobile traffic that varies in both temporal- and spatial-dimension. Due to its unique characteristics including vast coverage, fast deployment, high scalability, and easy programmability, UAV-assisted communication emerges as a promising approach to resolve the above scarcity of terrestrial networks \cite{AAGI,WCWU}. As illustrated in Fig.~\ref{UaaS_C1}~(a), a network of UAVs work as supplementary alternatives for those overloaded and malfunctioned BSs, to provide an ubiquitous wireless coverage over the serving area; UAVs can also be used as effective relays between two distant vehicles for reliable direct communication without the intervention of BSs, \textit{e.g.}, road safety information exchange.

Despite the attractive benefits by UAVs, it is nontrivial to realize those benefits in practice. For example, UAVs with on-demand mobility can help offload huge cellular data in some congested hot-spot areas at some specific time, \textit{e.g.}, shared bicycle parking lots during the rush hours. Due to the limited energy of UAVs, how to manage the UAV mobility and mission scheduling to cater for such a demand is challenging \cite{AGIM}. As pointed out in \cite{AGIM}, the information of the user mobility and service request is critical for UAVs to design the trajectory planning. We believe that in the UaaS architecture, the demand information can be predicted by advanced machine learning techniques (\textit{e.g.}, deep learning), according to the historical data including movement trajectory and request rate history. Based on that accurate information, UAVs intelligently choose their routine flying paths to meet the traffic demand in the congested areas at the right time.

%
%

\textbf{UAVs as an intelligent edge network regulator:} UAVs can also play an important role in the management of 6G networks. Due to their high altitude above the ground, UAVs are able to have line of sight (LoS) connection with multiple terrestrial users including both mobile users and ground BSs. Therefore, just as a traffic policeman standing at the crossroad has broad vision over multiple directions and is able to direct traffic, a UAV, like an ``aerial traffic policeman'', also has the potential to direct communication traffic and regulate the edge network behavior. As shown in Fig.~\ref{UaaS_C1}~(b), several UAVs cover four neighbouring small cells with common overlapping areas, where the four BSs are not directly interconnected by backhaul links. The resources available at these cells should be jointly optimized to improve the overall network performance and achieve load balancing. Thanks to the LoS link between the UAVs and users within the coverage area, all the information including users' QoE requirements and available resources at the BSs can be first aggregated at each BS, which is then uploaded to the UAVs for making joint decision.

Directing communication traffic may be not as easy as directing traffic at the crossroad. On one hand, the joint access and resource allocation problems at the edge network generally are large-scale, complex, and non-convex. Traditional optimization methodologies such as combinatorial optimization and convex optimization techniques may be ineffective for such sophisticated problems. On the other hand, in practice, much of the environment information including wireless channel and mobility pattern is statistical, which suits for performance analysis on average and offline optimization, rather than decision making in real time. It calls for model-free learning-based approaches. In the UaaS, leveraging DRL \cite{DRL,ADRL}, UAVs can provide intelligent real-time decision-making for each cell, without accurate and complete network information. Specifically, a network of UAVs act as a learning agent, which observe the state of the terrestrial networks including resource availability, QoE requirements, mobility velocity, node location, and link quality. The agent then determines its action (\textit{i.e.}, access decision and resource allocation) after the reward prediction by DNNs. The real reward such as throughput, latency, packet loss rate, and QoE is returned to the agent to adaptively adjust its sequential decisions. 

%

\section{UAVs as an Edge Computing Service}
Considering that many emerging applications such as extended reality (XR) and autonomous driving require in-network and distributed computation, it is critical to provide computing service for a wide range of IoT devices with limited computation capabilities. With the continuous cost reduction and increasing capabilities in UAVs, it is expected that UAV-enabled MEC networks will be widely implemented in 6G networks. However, how to intelligently employ UAVs as an edge computing service remains largely unsolved. 

\begin{figure*}[t]
\centering
\subfloat[UAVs as an edge computing platform]{\includegraphics[width=3.3in]{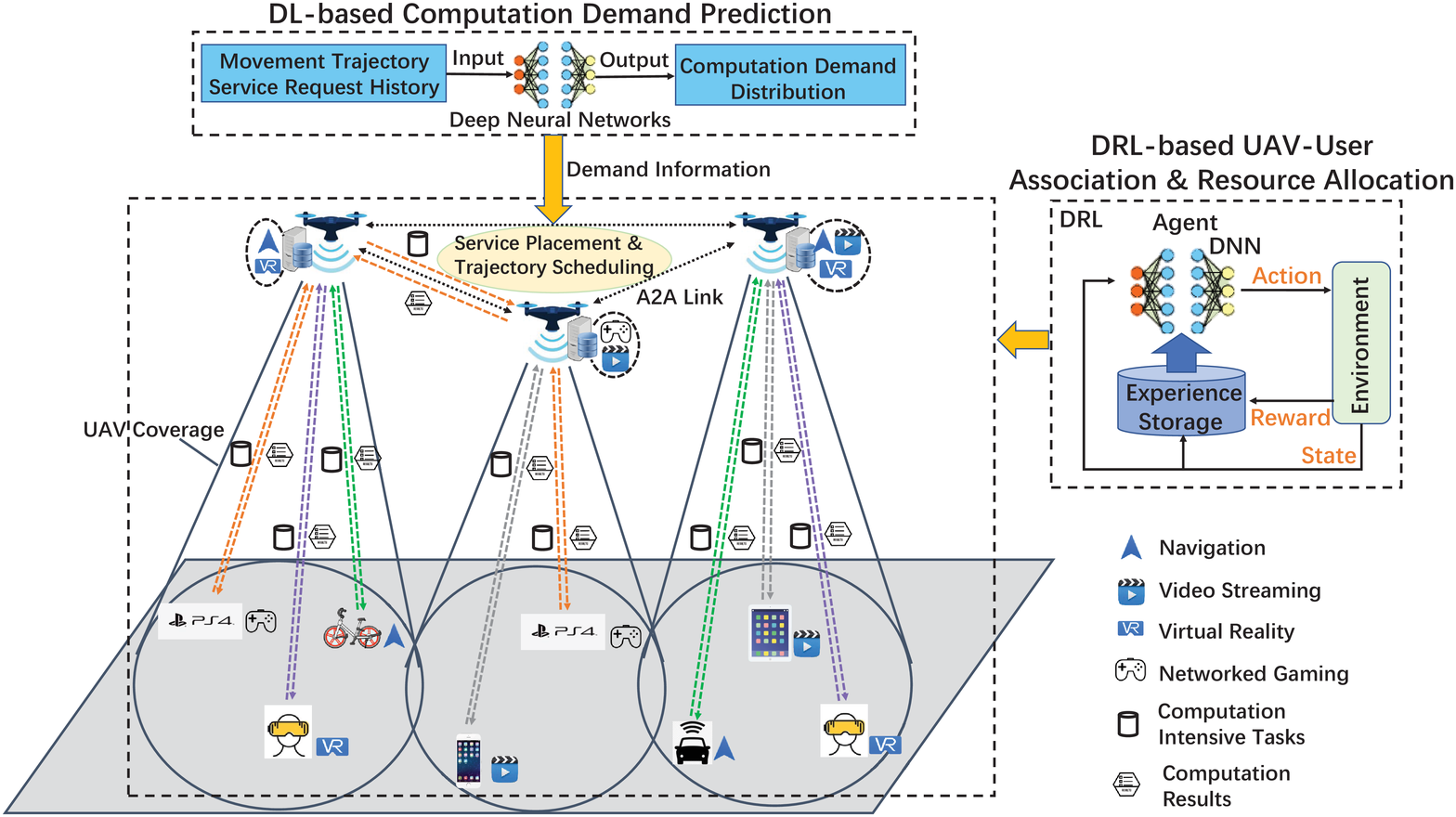}}\hspace{0.3cm}
\subfloat[UAVs as an effective edge intelligence trainer]{\includegraphics[width=2.8in]{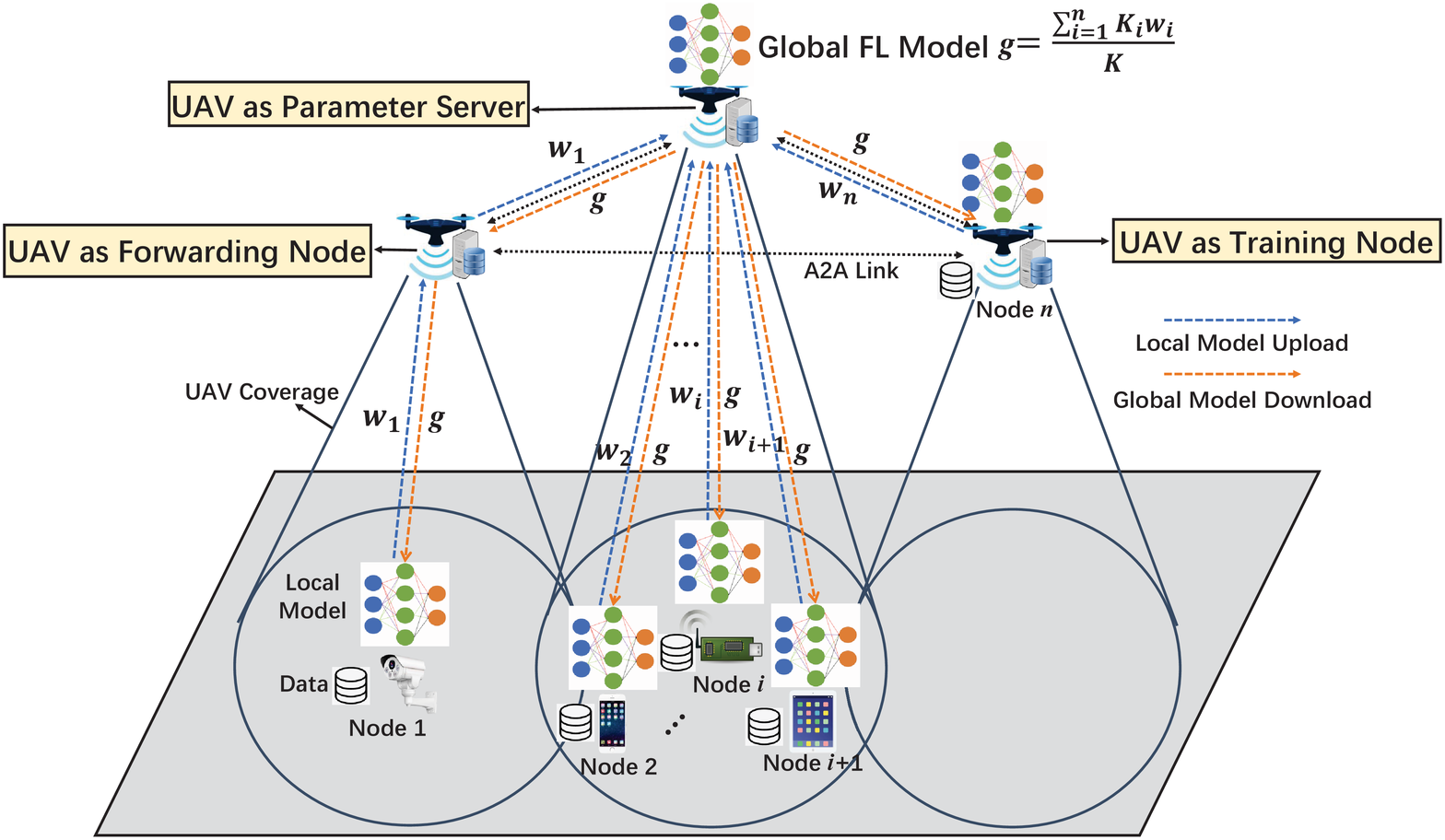}}\hfill
\caption{UAVs as an Edge Computing Service.}
\label{UaaS_C2}
\vspace{-0.5cm}
\end{figure*}

\textbf{UAVs as a smart edge computing platform:} UAVs with relatively powerful computing capability are used as on-demand mobile edge servers for terrestrial users. Unlike the case of UAVs as an ubiquitous aerial communication platform, the communications between them are computation-oriented in this case. Fig.~\ref{UaaS_C2} (a) illustrates the basic process of computation offloading in UAV-enable MEC networks. There are generally two offloading modes, \textit{i.e.}, binary and partial offloading, where in the former the whole computation task is offloaded to the UAV server for execution, while in the later only part of the task is offloaded. Zhou \textit{et al.} \cite{MECI} summarizes several prominent advantages of UAV-enabled MEC networks compared to the traditional terrestrial MEC networks, \textit{e.g.}, deployment flexibility, controllable mobility, and high availability in extreme situations such as emergency relief, disaster response, and military scenarios.

To be a smart edge computing platform, it is faced with several challenges as follows. First, to serve the computation request, accurately capturing the computation demand information inside the communication traffic is crucial. Furthermore, service provisioning, \textit{i.e.}, how to determine which service to be pre-stored at which UAV, to meet the diverse demands, is even more challenging. Last, how to efficiently allocate various resource at multiple UAV servers for multiple terrestrial users, as well as the trajectory scheduling of UAVs, is urgently needed. In the UaaS, we anticipate that, the computation demand information can be mined according to the knowledge gained from historical data by advanced ML techniques. That information is used to determine how many UAVs with how much computing and storage resources should be dispatched to cover the demand at the target area, while the service provisioning can be jointly determined simultaneously. For the joint resource allocation and trajectory scheduling in the multiple-user multiple-UAV scenario, DRL can also be used to solve such large-scale complex non-convex problem with highly dynamics.

\textbf{UAVs as an effective edge intelligence trainer:} UAVs can be employed to participate in the model training of distributed ML. This type of computing service does not need UAVs to have much powerful computation ability, as compared to the previous computing service. A large variety of edge users such as IoT devices are expected to gain AI insight, called as ``edge intelligence'', via making full use of the widespread edge resources, rather than totally relying on the cloud. Edge intelligence however, is provided that edge users themselves can effectively train the learning model. Federated learning (FL) \cite{FL}, as one kind of distributed ML, emerges as a promising technology to well preserve privacy, when training the DNN model according to the distributed local data. In the UaaS, we envision that UAVs can act as the central parameter server or a client in the FL model training, as shown in Fig.~\ref{UaaS_C2}~(b). A UAV as a parameter server is in charge of first collecting all the local model updates, calculating the global model, and then disseminating the global model to all training nodes, until the training converges. Additionally, UAVs can serve as flying training nodes. 

We believe that UAV-based FL model training has at least several outstanding advantages as follows. First, UAVs can be flexibly deployed to participate in a model training anywhere in theory. Second, UAVs have high probability of short-distance LoS links to every training client, which can greatly accelerate the training process. Third, a UAV is able to coverage a large number of users and recruit them into the training, without impacting the nearby cellular traffic. Last but not the least, UAVs with appropriate IoT devices flying in the sky may collect more abundant data including monitoring pictures and videos, compared to their terrestrial counterparts. Yet, it is highly desirable to design energy-efficient training strategies for UAV-based FL model training, since the energy resource of UAVs is limited and flying consumes plenty of energy, besides computation and communication.

\section{UAVs as an Edge Caching Service}
With the reduced cost and increased capacity of the storage devices, network caching strategies become practical to relieve the stress of the backhaul networks. Edge caching is considered as a promising approach to enhance users' QoE and provide backhaul and transport savings, by proactively caching popular contents at the network edge. How to realize that potential is of great importance to the objective of providing personal intelligent service in 6G networks. In this section, we envision that UAVs can play an important role in the edge caching for 6G, as proposed in the UaaS architecture. 

\begin{figure*}[t]
\centering
\subfloat[UAVs as an edge caching provider]{\includegraphics[height=1.8in]{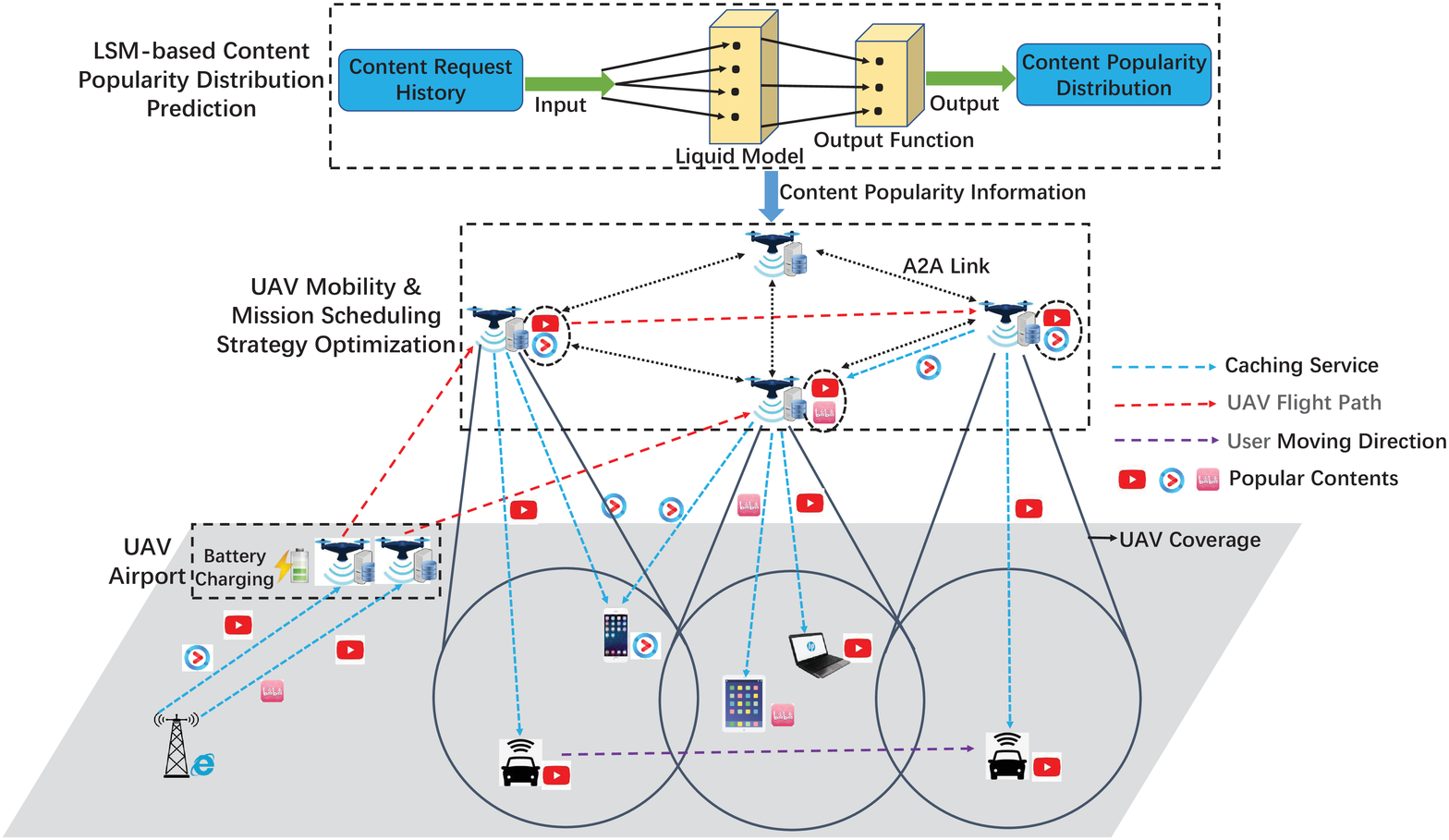}}\hspace{0.3cm}
\subfloat[UAVs as a clever edge content sharing assistant]{\includegraphics[height=1.8in]{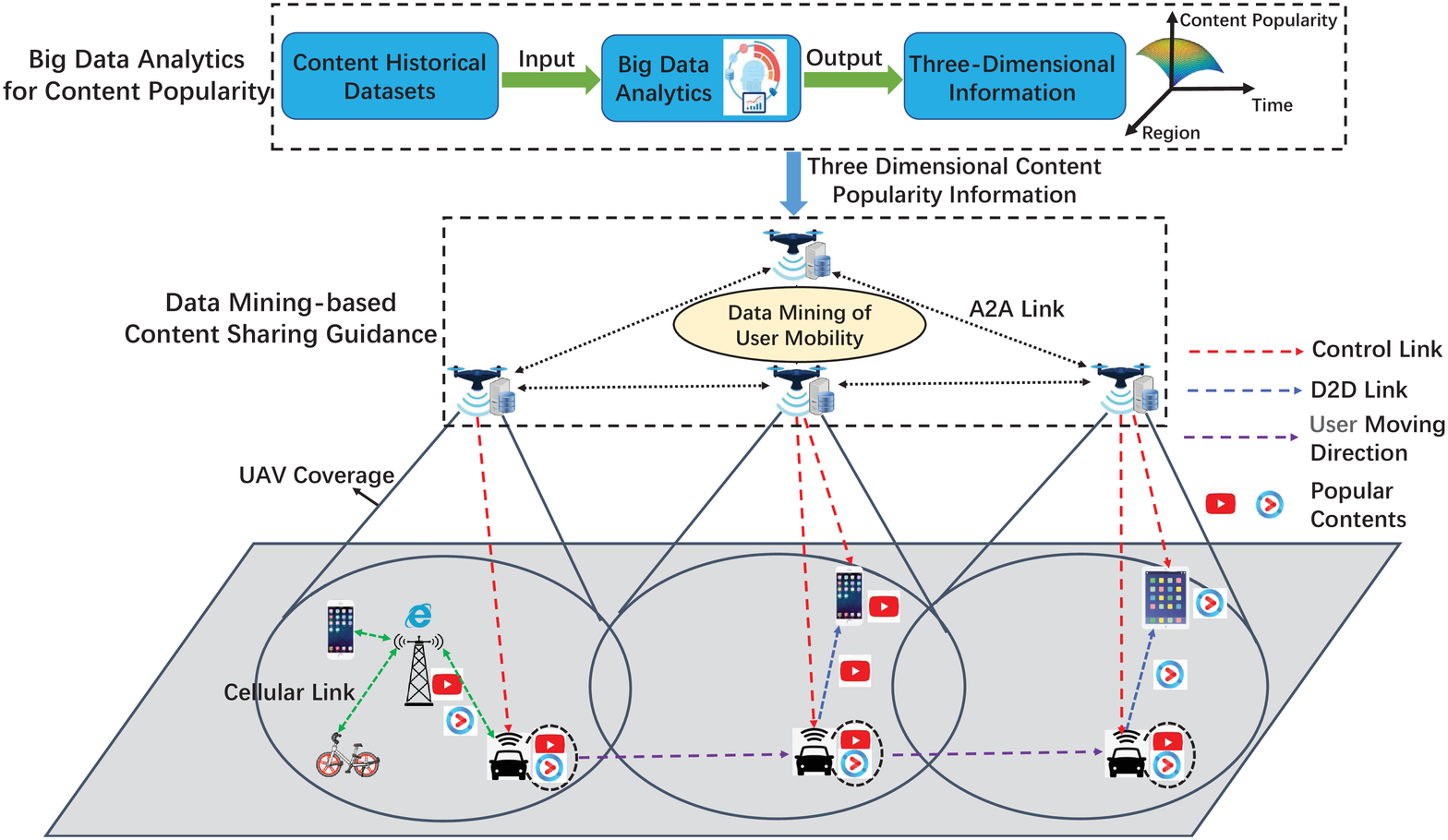}}\hfill
\caption{UAVs as an Edge Caching Service.}
\label{UaaS_C3}
\vspace{-0.5cm}
\end{figure*}

\textbf{UAVs as an intelligent content cache provider:} UAVs equipped with multiple cache storage units are deployed to serve the content request of terrestrial users. Specifically, some frequently requested contents, can be proactively cached in the UAVs, and further distributed to the requesting terrestrial users, without any further redundant transmissions in the backhaul. As shown in Fig.~\ref{UaaS_C3}~(a), the contents can be first downloaded and cached when UAVs stay at the docking station for battery charging. The UAVs are then dispatched to the target area, and serve the content requests from the sky. Note that the content caching can also be performed during off-peak hours. The first benefit of such a UAV-enabled content cache provider is that, both the service delay and backhaul burden can be reduced. The second one exists in the deployment flexibility. That is, due to the controllable mobility of UAVs, the repeatable content caching in different areas could be significantly reduced, since they can fly across the network to carry with popular contents, following user trajectories. Note that in the UaaS, there are normally multiple UAVs consisting of an integrated content cache provider, to cooperatively serve the content request in a large area.

To intelligentize the UAV-enabled content cache provider, we adopt the following measures step by step. First, the distribution of users' content requests can be predicted by ML techniques such as liquid state machine (LSM) \cite{LSM}. The content distribution should contain the timing information, \textit{i.e.}, which content is more popular in which time period. Second, based on the content popularity information, the content cache policy are jointly determined with the UAV mobility and mission scheduling, that is, which content should be cached at which UAV to serve which area. The above policy decision should also consider the current and future status of ground BSs in different areas. Last, the UAV-user association and corresponding resource allocation can be jointly optimized to maximize the caching performance. Since each UAV has been designated to serve an area, the above optimization problem within a single area may be non-convex but relatively not large, which can be efficiently solved by existing non-convex optimization techniques.

\textbf{UAVs as a clever edge content sharing assistant:} UAVs can assist to content sharing in terrestrial networks. It is similar to the aforementioned case of UAV as an intelligent edge network regulator, while content sharing without passing the cellular network as a special service is managed here. Specifically, following the schedule of UAVs, mobile users such as vehicles and mobile devices can cache the previously requested contents, and distribute these contents to the users they come across, via D2D transmissions. The prominent advantage of UAV-assisted edge sharing is that, by exploiting the wide coverage of UAVs, content sharing can be scheduled across multiple nearby cells/areas, such that the resources among these areas are efficiently utilized along with user mobility. As illustrated in Fig.~\ref{UaaS_C3}~(b), under the guidance of multiple UAVs, a user with high mobility is responsible for downloading a popular content from the cellular BS, and then shares it to the requested users in its route.

To be a clever edge sharing assistant, UAVs need to gather a lot information including users' content request in the dimensions of temporal and spatial besides the amount. And more critically, UAVs should have the ability of predicting the future popular content requests and hot areas, as well as the user mobility. In the UaaS, with big data analytics, UAVs are able to learn the three-dimensional information of which contents are popular during which time period at which area, based on histories. Additionally, according to the mining of user mobility, those users predicted to have high mobility across multiple areas are chosen to share the cached contents. Note that, UAVs should also make the real-time decision about the edge caching, in the light of current status.

\section{Case Study: UAV-enabled Federated Learning}
In this section, we conduct a case study to evaluate the performance of the proposed UaaS. Specifically, we focus on the scenario of UAVs as an effective edge intelligence trainer, where a UAV acts as the parameter server in the FL model training of multiple geographical located terrestrial users. To our best, this is the first work to discuss how to exploit UAVs for the effective model training of distributed ML. 
\vspace{-0.3cm}
\subsection{Evaluation Setting}
We consider a UAV-aided network consisting of one rotary-wing UAV having ability to hover and a set $\mathcal{N}$ of $N=100$ terrestrial user equipments (UEs), where each UE $n\in\mathcal{N}$ stores a local dataset. With the help of the UAV as the parameter server, these UEs cooperatively perform a FL model training for inference, where the training needs interactions between the UAV and UEs within multiple rounds. In each round, the UAV generates a global FL model and broadcasts it to the UEs, and then each UE updates its local FL model by it dataset and transmits it to the UAV. Following \cite{FL}, at the beginning of each training round, we only select a random fraction $\alpha=0.1$ of UEs for model update. The training task we consider here is image classification employing convolutional neural network (CNN) on the federated MNIST dataset \cite{FL}.

We assume that the computation capacity of UE $n\in\mathcal{N}$ and number of CPU cycles to execute one sample of data at this UE are randomly chosen from [1.8 GHz, 2.0 GHz] and fixed at 20 cycles/bit, respectively. For the UAV, we assume its propulsion power and transmission power equal to 100 W and 10 mW, respectively \cite{EERA}. For the communication between the UAV and UE $n\in\mathcal{N}$, the uplink data transmission rate is
\begin{align}
r_n=B_n\log_2\left(1+\frac{\beta_0 p_n}{\sigma^2(H^2+R^2_n)}\right),
\end{align}where $B_n$ is the bandwidth allocated to UE $n$, $p_n$ is the transmission power of UE $n$, $\beta_0$ is the channel power gain at the reference distance 1 m, $H$ is the UAV's flying height, $R_n$ is the horizontal distance between UE $n$ and the UAV, and $\sigma^2$ is the noise power. Following \cite{EERA}, the settings for the above parameters are as follows: $B_n=\frac{1}{\alpha N}B$ (the system bandwidth $B=1$ MHz), $p_n= 100$ mW, $\beta_0 = -50$ dB, $\sigma^2 = -110$ dBm, $H=100$ m, and $R_n$ randomly chosen from [0 m, 10 m]. Note that we employ the OFDMA access for the UEs with equal spectrum sharing in the above. The downlink data transmission rate can be calculated similar to Eq.~(1), except that $B_n$ and $p_n$ become 1 MHz and 10 mW corresponding to the UAV, respectively.
\begin{figure*}[t]
\centering
\subfloat[Testing accuracy \textit{vs.} training rounds]{\includegraphics[height=1.8in]{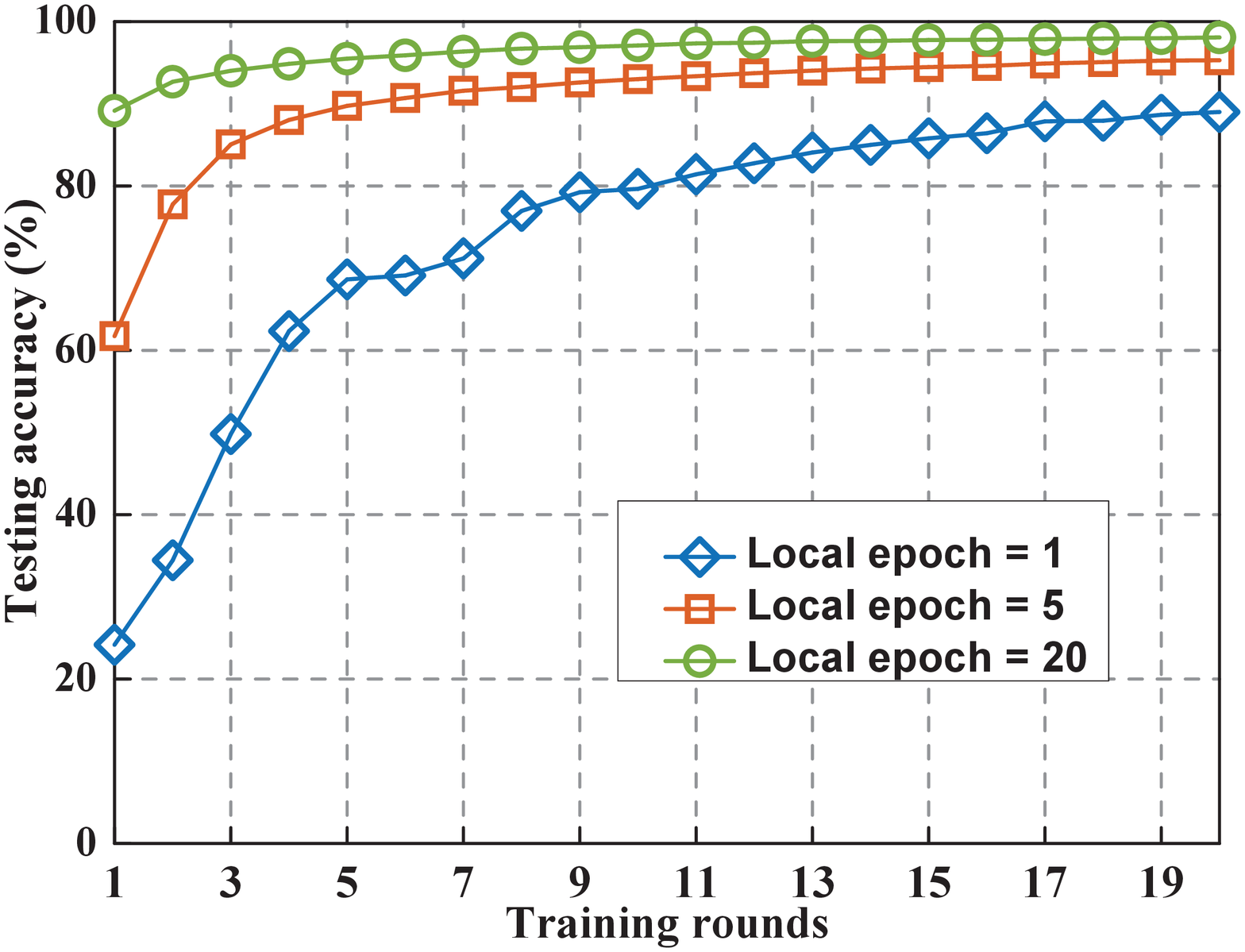}}\hspace{0.3cm}
\subfloat[Energy consumption \textit{vs.} training rounds]{\includegraphics[height=1.8in]{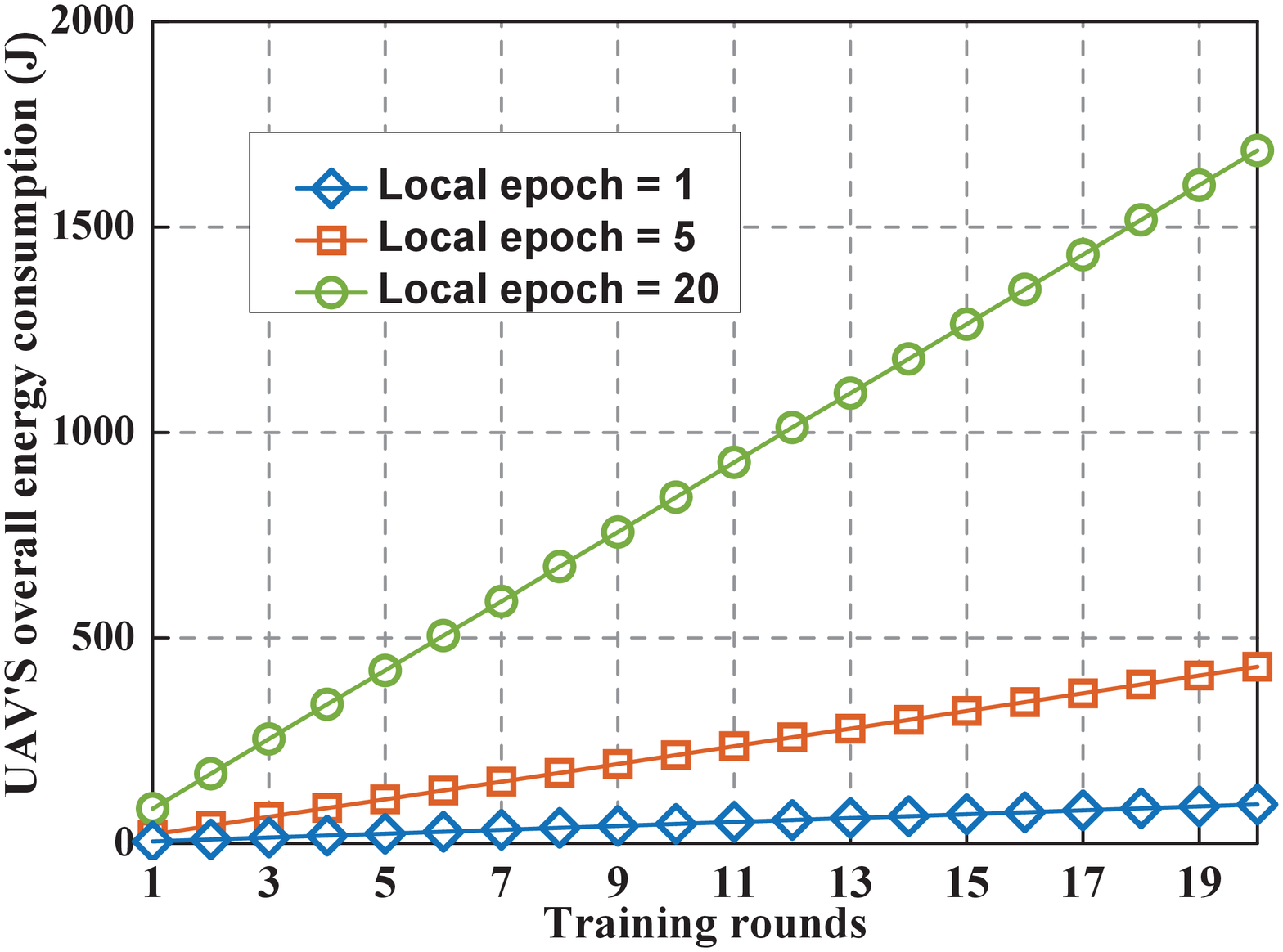}}\hfill
\caption{Performance of UAV-enabled FL model training.}
\label{UAV_FL}
\vspace{-0.5cm}
\end{figure*}
\vspace{-0.35cm}
\subsection{Results}
We evaluate the performance of the proposed UAV-enabled FL model training, in terms of testing accuracy and overall consumed energy of the UAV versus the number of training rounds. We also select three different number of local training epoches (\textit{i.e.}, 1, 5, 20) in the evaluation. The consumed energy of the UAV mainly consists of two parts: the energies to remain aloft and disseminate the global model update, which are calculated as the product of propulsion power and overall training time, and that of transmission power and overall global model update dissemination time, respectively. As shown in Fig.~\ref{UAV_FL}, we can achieve $\sim$97\% testing accuracy by consuming no more than 500 J energy of the UAV with only five local epoches. And our results also show that the total energy consumption for the global model dissemination equals to $\sim$0.0013112 J only, which can be neglected compared with the relatively huge flight energy consumption. This validates the effectiveness of the proposed UAV-enabled FL model training, which also shows its bright future.

\section{Challenges and Open Research Issues}
The proposed UaaS architecture envisions the various services that UAVs can intelligently provide for future 6G networks. However, there is still a long way to go before realizing the provisioned services by UAVs, where many key challenges and open research issues need to be addressed. 

\textbf{Flexible switching among communication, computing, and caching services:} in the previous sections, we introduce how UAVs can be as a wireless communication service, edge computing service, and edge caching service, respectively. In practice, these services should be integrated in any UAV, to provision service on demand. It calls for flexible switching among communication, computing, and caching under UaaS, which remains unsolved so far.

\textbf{Collaboration among a network of UAVs:} a single UAV may be unable to well provide the service in the envisioned UaaS architecture. As a matter of fact, to satisfy the huge demand of a variety of terrestrial users, the collaboration of multiple UAVs is indispensable. Furthermore, multiple UAVs should be intelligently controlled to provision a service as a whole. Exploiting multiple-agent reinforcement learning for the intelligent control of multiple UAVs is promising but also challenging \cite{MECI}.

\textbf{Efficient machine learning model training:} intelligence is the biggest characteristic of the proposed UaaS, which requires UAVs to gain intelligence by ML techniques. Due to the scarce resource and heavy responsibilities, it is urgent to perform energy-efficient machine learning model training. In particular, when UAV acts as an edge intelligence trainer, we should design energy-efficient training strategies for all participants, especially the participating UAVs.

\section{Conclusion}
In this article, we have proposed the UaaS architecture to boost edge intelligence for the air-ground integrated networks. Specifically, the UaaS employs UAVs to intelligently provision the three fundamental services in terms of wireless communication, edge computing, and edge caching, with the help of advanced ML techniques. We have described the architecture, envisions, challenges, and solutions of UaaS, introduced how to realize each service in detail, and conducted a case study to validate the effectiveness of UaaS in boosting edge intelligence. Several key challenges and promising research issues are also highlighted for future study.

\bibliographystyle{IEEEtran}

\end{document}